\documentclass[prd,preprintnumbers,superscriptaddress,nofootinbib]{revtex4}
\usepackage{amsmath}
\usepackage{amssymb}
\usepackage{hyperref}
\usepackage{float}
\usepackage{xspace}
\usepackage{slashed}
\hypersetup{
    colorlinks=true,       % false: boxed links; true: colored links
    linkcolor=blue,          % color of internal links
    citecolor=blue,        % color of links to bibliography
    filecolor=blue,      % color of file links
    urlcolor=blue           % color of external links
}
\usepackage{xcolor}
\usepackage{color}
\usepackage{graphicx}  %
\usepackage{bm}  %                                          
\usepackage{amsmath}   %
\usepackage{graphics}%, slashed, braket}
\usepackage{ulem}
\usepackage{comment}

\newcommand{\dslash}[1]{#1 \llap{/\kern-0.5pt}}
\newcommand{\Dslash}[1]{#1 \llap{/\kern+1.5pt}}
\newcommand{\DDslash}[1]{#1 \llap{/\kern+2.3pt}}
\newcommand{\dslashh}[1]{#1 \llap{/\kern+1pt}}

\newcommand{\bea}{\begin{eqnarray}}
\newcommand{\eea}{\end{eqnarray}}
\newcommand{\be}{\begin{equation}}
\newcommand{\ee}{\end{equation}}
\newcommand{\bma}{\begin{pmatrix}}
\newcommand{\ema}{\end{pmatrix}}

\allowdisplaybreaks[1]

\def\Cll{\lbrack C_{\ell\ell} \rbrack}
\def\Cphil{\lbrack C_{\phi\ell}^{(3)}\rbrack}
\def\Cphiq{\lbrack C_{\phi q}^{(3)} \rbrack}
\def\Clq{\lbrack C_{\ell q}^{(3)} \rbrack}
\def\Clqone{\lbrack C_{\ell q}^{(1)} \rbrack}

\def\epsll{\lbrack\epsilon_{\ell\ell}\rbrack}
\def\epsphil{\lbrack\epsilon_{\phi\ell}^{(3)} \rbrack}
\def\epsphiq{\lbrack \epsilon_{\phi q}^{(3)} \rbrack}
\def\epslq{\lbrack \epsilon_{\ell q}^{(3)} \rbrack}

\def\Oll{\lbrack O_{\ell\ell} \rbrack}
\def\Ophil{\lbrack O_{\phi\ell}^{(3)} \rbrack}

\def\Ophiq{\lbrack O_{\phi q}^{(3)} \rbrack}
\def\Olq{\lbrack O_{\ell q}^{(3)} \rbrack}
\def\Olqone{\lbrack O_{\ell q}^{(1)} \rbrack}

\begin{document} 

\preprint{
TIFR/TH/21-11
}

\preprint{{LAPTH-028/21}
}

\title{
%{Leptonic Operators and RG Running for Cabbibo Angle Anomaly}
%}

{Leptonic Operators for Cabbibo Angle Anomaly with SMEFT RG Evolution}
}

\author{Ashutosh Kumar Alok}
\email{akalok@iitj.ac.in}
\affiliation{Indian Institute of Technology Jodhpur, Jodhpur 342037, India}

\author{Amol Dighe}
\email{amol@theory.tifr.res.in}
\affiliation{Tata Institute of Fundamental Research, Homi Bhabha Road,  Colaba, Mumbai 400005, India}

\author{Shireen Gangal}
\email{gangal@lapth.cnrs.fr}
\affiliation{LAPTh, Universite Savoie Mont-Blanc et CNRS, Annecy, France}

\author{Jacky Kumar}
\email{jacky.kumar@tum.de}
\affiliation{Institute for Advanced Study, Technical University Munich, 85748 Garching, Germany}

\begin{abstract}
The measurements of the 
Cabibbo--Kobayashi--Maskawa (CKM) elements can be 
contaminated by new-physics effects. 
We point out that purely leptonic operators at the high scale can influence 
semileptonic $K$ decays and nuclear beta decay through renormalization group (RG) running, 
and hence can influence the measurements of $V_{us}$. 
Interestingly, through this mechanism, a single six-dimensional effective operator {$O_{\ell\ell}$}
 at the high scale can {alleviate the tension due to} the Cabibbo angle anomaly, by
generating the desired operators at the low scale through RG 
running.  When generated as a result of a $Z'$ model, the non-universal leptonic
couplings of this operator can also contribute to the lepton
flavor universality violating ratios such as $R_{K^{(*)}}$, which would  act as 
stringent constraints on such scenarios. 
{By performing a global fit of the $Z'$ model, we find that} it is essential to have non-universal 
couplings of such a $Z'$ boson to all three generations of leptons.
\end{abstract}

\maketitle

\section{Introduction}
The standard model (SM) of particle physics encodes our current understanding
of fundamental interactions in nature. Since the advent of this theory in the
mid-1970s, a large number of experiments have tested its several aspects.
The SM has successfully accounted for most of the experimental measurements
within its domain, giving us confidence in its foundations.  
However, it cannot be a complete theory, as it fails to explain the
observed baryon asymmetry in the Universe, the nature of dark matter
and dark energy, and gravitational interactions.
The exploration of physics beyond SM is carried out via two modes --
direct searches where new heavy particles may be produced at high-energy
particle colliders, and indirect searches, where the effects of these heavy
particles may be detected through the quantum corrections they give rise to,
even at energies lower than their masses. The latter is the preferred mode of
operation of flavor physics, wherein precision measurements can probe for
effects of particles much heavier than energies accessible at present-day
colliders.

In the absence of any concrete clue about the kind of new physics (NP)
at high energies, one may use the Standard Model Effective Field Theory
(SMEFT) framework, where the SM is extended with a series of higher-dimensional
operators $O_i$, while keeping its gauge symmetries intact~ \cite{Buchmuller:1985jz, Grzadkowski:2010es}.
This allows the introduction of NP in a model-agnostic way.
Limiting ourself to dimension-six operators, one may write the SMEFT 
Lagrangian as 
\begin{equation}
  \mathcal{L}_{\rm eff}^{\rm SMEFT} =  \mathcal{L}_{\rm SM} + \sum_{i} {C_i O_i} + ...
  \; .
  \label{eq:leff}
\end{equation}
Here, the $C_i$'s are known as Wilson coefficients (WCs) that can be calculated
perturbatively. Note that the WCs
are scale dependent quantities, whose values at a given scale may be
calculated using renormalization group running
equations~\cite{Alonso:2013hga, Jenkins:2013wua}.
In our analysis, we use the Warsaw-down basis in the WCxf conventions~\cite{Aebischer:2017ugx}.
  
One of the precision observables that has shown signs of NP is the
measurement of the element $V_{us}$ of the Cabibbo-Kobayashi-Maskawa (CKM)
matrix which describes the mixing of quarks. The measurement of this
quantity (also called the Cabibbo angle) from different processes
like  
nuclear beta decay~\cite{Seng:2018yzq,Seng:2018qru,Czarnecki:2019mwq, Seng:2020wjq, Hayen:2020cxh, Shiells:2020fqp},
Kaon decay~\cite{Antonelli:2010yf, Moulson, Seng:2019lxf, Seng:2020jtz, Seng:2021boy, Seng:2021wcf, Seng:2021nar}, 
tau decay~\cite{Amhis:2019ckw}, and the global fit~\cite{Grossman:2019bzp} to
all elements of the CKM matrix give slightly incompatible values.
This discrepancy is known as the ``Cabibbo Angle Anomaly" (CAA).

The element $|V_{us}|$ can be determined from semileptonic Kaon decays
  $K \to \pi \ell \nu$ ($K_{\ell 3}$), where $\ell$ is either an electron or muon. 
    Using the vector form factor at zero momentum $f_+(0)$ from lattice QCD with
  $N_f =2+1+1$ flavors \cite{FermilabLattice:2018zqv}, one gets, 
$|V_{us}^{K_{\ell 3}}|= 0.22306 \pm 0.00056$ \cite{Seng:2021nar}. The ratio of decay rates of $K \to \mu \nu(\gamma)$
  and $\pi \to \mu \nu(\gamma)$ can be used to determine $|V_{us}/V_{ud}|$, using the lattice QCD
results for the decay constants, $f_{K}/f_{\pi}$. The value of this ratio is determined to be 
$|V_{us}/V_{ud}|= 0.23131 \pm 0.00051$ \cite{Seng:2021nar} which gives 
$|V_{us}^{K/\pi}| = 0.2252 \pm 0.0004$.  

Another way of determining $|V_{us}|$ is through the CKM unitarity relation $|V_{ud}|^2 +
|V_{us}|^2 \approx 1.0000$ and the measurement of $V_{ud}$.
The determination of $|V_{ud}|$ from super-allowed $\beta$ decays involves corrections due to nuclear structure
and nucleus independent electroweak radiative effects ($\Delta_R^V$). Over the last few years, there has been significant progress in the determination of $\Delta_R^V$ which involve calculations of $\gamma W$ box diagrams
using different approaches. 
Calculations by three groups -- Seng, Gorchtein, Patel, Ramsey-Musolf (SGRM)  \cite{Seng:2018yzq,Seng:2018qru},  Czarnecki, Marciano, Sirlin (CMS) \cite{Czarnecki:2019mwq}
and Shiells, Bluden, Melnitchouk (SBM) \cite{Shiells:2020fqp} -- lead to slightly different results:
 $|V_{ud}|_{\rm SGRM} = 0.97369 \pm 0.00014$, 
$|V_{ud}|_{\rm CMS} = 0.97389 \pm 0.00018$ and $|V_{ud}|_{\rm SBM} = 0.97368 \pm 0.00013$.
 Using unitarity,  this leads to 
$|V_{ us}^{\beta}|_{\rm SGRM} =0.22782 \pm 0.00062$, $|V_{us}^{\beta}|_{\rm CMS} =0.22699 \pm 0.00078$ and $|V_{ us}^{\beta}|_{\rm SBM} =0.22782 \pm 0.00062$.
Further nuclear corrections in $0^+ \to 0^+$ transitions \cite{Gorchtein:2018fxl} would leave the central values of $|V_{us}^{\beta}|$ unchanged, but would increase the uncertainties. 

Inclusive and exclusive $\tau$ decays can also be used to determine $|V_{us}|$. 
 Inclusive $\tau$ decays to final states involving
  strange quarks give $|V_{us}^{\tau}| = 0.2195 \pm 0.0019$ \cite{Amhis:2019ckw}.
  This extraction of $|V_{us}|$ depends upon the calculation of corrections due to finite quark masses
  and non-perturbative QCD effects \cite{Gamiz:2004ar,Gamiz:2002nu}. 
  The determination of $|V_{us}^{\tau}|$ from the ratio of decay rates
  $\Gamma(\tau \to K \nu)/\Gamma(\tau \to \pi \nu)$ is $0.2236 \pm 0.0015$,
  while that from $\tau \to K \nu$ decays is  $0.2234 \pm 0.0015 $ \cite{Amhis:2019ckw}.

  It is evident that the above measurements of $|V_{us}|$ from different decay modes are incompatible with each other.
  Compared to the CKM unitarity prediction of $0.2245 \pm 0.0008$
  \cite{pdg}, the $|V_{us}^{\tau}|$ value from the inclusive $\tau$ decays
  is smaller by $\sim 2.9\sigma$, while the average from inclusive and exclusive $\tau$ decays, $|V_{us}^{\tau}| = 0.2221 \pm 0.0013$
   is smaller by $\sim 2\sigma$ \cite{pdg}.
  The $\beta$ decay measurements, on the other hand, yield $|V_{us}^\beta|$ values
  that are higher than the unitarity prediction, the level of inconsistency
  depending upon the radiative corrections scheme. 
Using the latest prediction of $|V_{ud}| = 0.9737 \pm 0.00030 $ which includes the nuclear structure 
uncertainties \cite{Hardy:2020qwl}, 
the unitarity relation gives $|V_{ud}|^2 + |V_{us}|^2 - 1 = -0.0021 \pm 0.0006$, which indicates an apparent anomaly in the top row CKM 
unitarity at the level of $3.2\sigma$ \cite{Seng:2021nar}.

The CAA may be quantified through the measurement of the ratio
\be
R(V_{us}) \equiv  |V_{us}^K| / |V_{us}^\beta|  \; ,
\ee
where $|V_{us}^K|$ is the value obtained from semileptonic decays of $K$,
while $|V_{us}^\beta|$ is the value obtained from nuclear beta decays and
the unitarity relation $|V_{ud}|^2 + |V_{us}|^2 \approx 1.0000$. The measured value
of this ratio is~\cite{Crivellin:2020lzu}
\be
R(V_{us}) = 0.9891 \pm 0.0033 \;,
\ee
which is more than $3\sigma$ away from the expected value of unity.

The CAA has been interpreted as a possible sign
for the violation of the CKM unitarity~\cite{Belfatto:2019swo,Cheung:2020vqm,
Felkl:2021qdn,Belfatto:2021jhf,Branco:2021vhs}, which is one of the {pillars}
of the SM. However, it can also be resolved keeping the CKM unitarity intact,
provided lepton flavor universality (LFU) violating NP couplings of $W$ bosons
to leptons are invoked~\cite{Coutinho:2019aiy,Crivellin:2020lzu}.
The latter resolution, in its simplest form, is in tension with the electroweak
precision (EWP) {observables}~\cite{Kirk:2020wdk}, since the $SU(2)_L$
symmetry of SM also mandates NP couplings to the $Z$ boson.
The most natural way to alleviate this tension is to have additional sources
of gauge-invariant couplings of the $Z$ boson to the left-handed
leptons~\cite{Alok:2020jod}. The connection between CAA and
other observables has been studied in 
Refs.~\cite{Crivellin:2021bkd, Crivellin:2021rbf, Crivellin:2021njn, Crivellin:2020klg}.

A measurement of the ratio ${\rm BR}(K \to \pi \mu \bar \nu)/{\rm BR}(K \to \mu \bar \nu)$, 
possible at the NA62 experiment, can help to determine whether the current tensions are due 
to possible physics beyond the SM or experimental issues \cite{Cirigliano:2022yyo}. 
Future improvements in the calculations of nuclear corrections can also impact the extent of 
CAA \cite{Hardy:2020qwl, Gorchtein:2018fxl, Seng:2022inj}.  

In this work, we address the CAA in the SMEFT framework, specifically focusing on the 
pure leptonic operators at the NP scale. We systematically study the impact of 
these operators on CAA through the SMEFT renormalization-group running effects. 
As an example, we also study models involving a $Z'$ boson. With non-universal leptonic
couplings, a $Z'$ can give rise to leptonic SMEFT operators at 
the NP scale after it has been integrated out. 
 Such a $Z'$ model having minimal couplings to the leptons, 
bottom and strange quarks is well known to be able to address the
$B$ anomalies \cite{Alok:2017sui, DiChiara:2017cjq}. Therefore, $Z'$ models have potential to 
address the CAA and $B$-anomalies simultaneously\footnote{Note that the latest LHCb results suggest  
that the lepton flavor universality violating observables $R_{K^{(*)}}$ \cite{LHCb:2022zom} 
are consistent with the SM. However, the other $B$-anomalies in the branching 
fractions and angular observables still exist 
\cite{Alguero:2023jeh}.}. 

This work is organized as follows. In sec.~\ref{caa}, we use the effective field theory 
language and derive a general expression for the observable $R(V_{us})$  
in terms of SMEFT operators at the electroweak scale. We also study how pure 
leptonic operators can generate the operators that contribute to 
 $R(V_{us})$ through RG running effects. 
In sec.~\ref{Zp}, we show that the model with a $Z'$ boson is a viable candidate
for such an explanation, and that such a model may also be able to
account for $b \to s \mu^+ \mu^-$  data at the same time. We present constraints
from experimental measurements on such a generic 
$Z'$ model and present our fit results in sec.~\ref{results}.
We summarize our findings in sec.~\ref{summary}.

\section{Cabibbo angle anomaly in SMEFT}   
\label{caa}

The determination of $R(V_{us})$ depends on the measurements of $K$ decay and
nuclear $\beta$ decay. The six-dimensional SMEFT operators that are relevant
for these measurements are
\bea
\Oll_{ijmn} & \equiv & (\bar\ell_i \gamma^\mu \ell_j)
(\bar\ell_m \gamma_\mu \ell_n) \; , 
\label{op1}\\
\Olq_{ijmn} & \equiv & (\bar\ell_i \gamma_\mu \ell_j)
(\bar{q}_m \gamma^\mu q_n) \; ,
\label{op2} \\
\Ophiq_{mn} & \equiv & (\phi^\dagger i {\overleftrightarrow{D}_\mu^I} \, \phi)
(\bar q_m \tau^I \gamma^\mu q_n) \; , 
\label{op3}\\
\Ophil_{mn} & \equiv & (\phi^\dagger i {\overleftrightarrow{D}_\mu^I} \, \phi)
(\bar\ell_m \tau^I \gamma^\mu \ell_n) \; . 
\label{op4}
\eea
Here, $i, j, m, n$ are fermion generation indices.
The corresponding Wilson coefficients
are $\Cll_{ijmn}$, $\Clq_{ijmn}$, $\Cphiq_{mn}$, and
$\Cphil_{mn}$, respectively, and the relevant dimensionless parameters are
defined as $\lbrack \epsilon \rbrack = v^2 \lbrack C \rbrack$.
We take all WCs to be real, for the sake of simplicity.

In the presence of NP, the measured value of $R(V_{us})$
may be written as
\begin{equation}
  R(V_{us}) = 1 + \epsilon^{(0)} + \frac{\epsilon^{(1)}}{\lambda}
  + \frac{\epsilon^{(2)}}{\lambda^2} \; ,
  \label{eq:RVus-expansion}
\end{equation}
where $\lambda \equiv V_{us}/ V_{ud}$, and
\bea
\epsilon^{(0)} & = & -\epsphil_{11} 
+ \epsphiq_{22} - \epslq_{2222} + \frac{1}{2}\epsll_{1221} ,\\
\epsilon^{(1)} & = & \epsphiq_{21} + \epsphiq_{12}
- \epslq_{2212} - \epslq_{1121} , \\
\epsilon^{(2)} & = & -\epsphil_{22} 
+ \epsphiq_{11} - \epslq_{1111} + \frac{1}{2} \epsll_{1221}.
\eea
Here, the $\epsilon^{(1)}$ term is enhanced by a single power of
$(1/\lambda) \approx 5$, and the $\epsilon^{(2)}$ term is enhanced by 
$(1/\lambda)^2 \approx 25$, as compared to $\epsilon^{(0)}$. It is obvious 
that in general the effect on $R(V_{us})$  is not only through 
the modification of the Fermi constant $G_F$ which would come from $\epsll_{1221}$  
and $\epsphil_{22}$, but also from the other quantities, {\it viz.} $\epsphiq_{21}$, $\epslq_{2222}$, 
$\epslq_{1111}$, $\epslq_{2212}$  and $ \epslq_{1121}$.

We consider a situation where all NP WCs are zero at a high scale $\Lambda$,
except for $\Cll_{1111}, \Cll_{2222}$ and $\Cll_{1122}$.
This scenario is possible if a new particle couples with the first two
generations of leptons with diagonal couplings in the flavor basis.
Below the scale $\Lambda$, renormalization group (RG) evolution would
generate new operators of the type $\Ophil, \Ophiq, \Olq$, as well as
other elements of $\Oll$.
With the boundary conditions described above, the RG equations
\cite{Alonso:2013hga}, at the leading order, are  
\bea
16\pi^2 \frac{\mu \, d \epsilon^{(0,2)}}{d\mu} & \approx &
6 g_2^2 \epsll_{1122} \; ,
\label{eq:RG-eps2}\\
16\pi^2 \frac{\mu \, d \epsilon^{(1)}}{d\mu} & \approx  & 0 \; .
\label{eq:RG-eps1}
\eea
Since $\epsilon^{(0,1,2)}$ themselves are zero at the scale $\Lambda$,
this ensures that $\epsilon^{(1)}$ does not get produced by RG evolution,
and $\epsilon^{(0)}(\mu) = \epsilon^{(2)}(\mu)$.
The value of $R(V_{us})$, which is unity at the high scale, becomes 
\be
R(V_{us}) \approx  1  + \left[ 1 + \left( \frac{V_{ud}}{V_{us}} \right)^2\right]
\epsilon^{(2)}(\mu_{\rm EW}) 
\ee
at the low scale $\mu_{\rm EW}$. In the
leading log-approximation, the solutions to eq.~(\ref{eq:RG-eps2}) give
\be
\epsilon^{(2)} (\mu_{\rm EW})
\approx   -\frac{3 g_2^2}{8 \pi^2} \epsll_{1122}
\log \left( \frac{\Lambda}{\mu_{\rm EW}} \right) \; .
\ee
The deviation of $R(V_{us})$ from unity may be accounted for
by a non-zero value of $[\epsilon_{\ell\ell}]_{1122}$
corresponding to 
  \be \label{eq:bfeft}
   \Cll_{1122}(\Lambda)  =  0.47 \pm 0.14 ~\mbox{TeV}^{-2} \,,
  \ee
where we have taken $\Lambda = 1$ TeV and $\mu_{\rm EW} ~\simeq~ 91$ GeV.
{This value of $\Cll_{1122}$ is found to be consistent with the LEP constraints \cite{ALEPH:2013dgf} 
within $2\sigma$, even though the best fit point may be disfavored.}

Note that the WCs $\Cll_{1111}$ and $\Cll_{2222}$ have played no part in the
above, given our analytic approximations.
So in principle, the presence of only nonzero $\Cll_{1122}$ of an
appropriate value at the high scale $\Lambda$ is sufficient for generating
R$(V_{us})$. Thus, this is a one-parameter solution for
resolving the CAA.

%%%%%%%%%%%%%%%%%%%%%%%%%%%%%%%%%%%%%%%%%%%%%%%%%%%%%%%%%%%%%%
\begin{figure}[bht!]
  \includegraphics[width=0.45\textwidth]{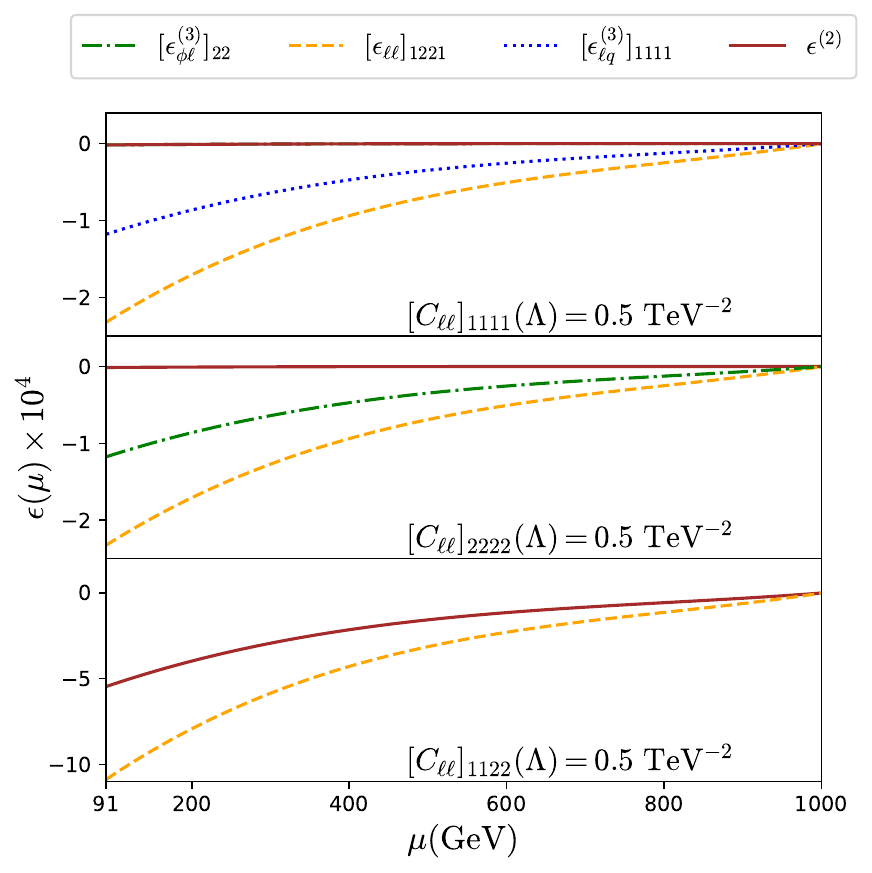}
  \caption{The RG evolution of the effective NP parameter
    $\epsilon^{(2)}(\mu)$, and terms contributing to it.
    The top, middle, and bottom panels correspond to the scenarios
    where $\Cll_{1111}(\Lambda)$, $\Cll_{2222}(\Lambda)$, 
    and $\Cll_{1122}(\Lambda)$ are nonzero, respectively.
    \label{fig:RGE}}
\end{figure}
%%%%%%%%%%%%%%%%%%%%%%%%%%%%%%%%%%%%%%%%%%%%%%%%%%%%%%%%%%%%%%

We confirm our analytic solution, and the negligible effect of
approximations employed therein, by solving the relevant sets of
RG evolution equations~\cite{Alonso:2013hga} numerically using
the {\tt wilson} package~\cite{Aebischer:2018bkb}.
The RG evolutions of terms contributing to $\epsilon^{(2)}(\mu)$
are shown in fig.~\ref{fig:RGE}.
From this figure, it is evident that there is no net effect of
$\Cll_{1111}(\Lambda)$ and $\Cll_{2222}(\Lambda)$
on the NP parameter $\epsilon^{(2)}(\mu)$.
Indeed, their effects on the component terms are seen to cancel\footnote{Similar cancellations 
also take place in the 1-loop SMEFT contributions to other electroweak parameters~\cite{Kumar:2021yod}.}.
On the other hand, nonzero $\Cll_{1122}(\Lambda)$ gives rise to
nonzero $\epsilon^{(2)}(\mu)$, and hence can account for
$R(V_{us})$.
This indicates that the resolution of the CAA necessarily requires NP in
the electron  as well as muon sector.
This is contrary to the earlier solutions proposed in terms of the operator
$\Ophil$, in which NP only in the muon sector was indicated
~\cite{Crivellin:2020ebi, Kirk:2020wdk, Endo:2020tkb, Alok:2020jod}.

One important prediction of this scenario is a shift in the value of the bare Fermi constant due to  
non-zero value of $[C_{\ell\ell}]_{1221}$. In SMEFT,
 \footnote{It is worth reminding that the
$\Cll_{2112}$ contribution is omitted as compared to Ref.~\cite{Alonso:2013hga}
since we are in the non-redundant flavor basis.} at the EW scale we have  \cite{Alonso:2013hga}
\begin{equation} \label{eq:dGFsmeft}
\frac{\delta G_F}{G_F^{(0)}} = v^2 \left  ( - \frac{1}{2} \Cll_{1221}(\mu_{\rm EW}) 
+  {[C_{H \ell }^{(3)}]}_{11}(\mu_{\rm EW}) + {[C_{H \ell}^{(3)}]}_{22}(\mu_{\rm EW} )  \right )\,,
\end{equation}
where the $\delta G_F$ can be defined through effective Fermi constant in SMEFT 
\begin{equation} \label{eq:GFsmeft}
G_F^{\rm SMEFT}= G_F^{(0)} \left ( 1 + \frac{\delta G_F}{G_F^{(0)}} \right ) \,,
\end{equation}
and we have defined the bare Fermi constant to be 
$G_F^{(0)} = 1/(\sqrt{2} v^2 )$.
In the definition of $G_F^{\rm SMEFT}$ through the eqs. \eqref{eq:dGFsmeft}-\eqref{eq:GFsmeft}, 
we have neglected the higher order SMEFT power corrections due to 
dimension-six contributions to vacuum expectation value ($v$).
At the best-fit point in Eq.~\eqref{eq:bfeft}, we obtain $\delta G_F/G_F^{(0)} \approx 5\times 10^{-4}$.
{Thus, our SMEFT scenario predicts\footnote{The $G_F^{\rm SMEFT} =1.1664\times 10^{-5} {\rm GeV^{-2}}$ in SMEFT can be extracted
through muon decay. Whereas the WCs $ \Cll_{1221}(\mu_{\rm EW}) $ is fixed by $R(V_{us})$ and a combination of these two 
provides us $G_F^{(0)}$ within SMEFT as given by Eq.\eqref{eq:GFsmeft}. On the other hand in the SM $G_F^{(0)} = 1.1664\times 10^{-5} {\rm GeV^{-2}}$ can be extracted solely from muon decay.
} that the value of the bare Fermi constant $G_F^{(0)}$, as determined through $R(V_{us})$, 
is less by $0.05\%$ than that measured through the muon decay. That is, in SMEFFT  $G_F^{(0)} = 1/(\sqrt{2} 
v^2)=1.1659\times 10^{-5} {\rm GeV^{-2}}$, whereas  $G_F^{(\mu)}= 1.1664\times 10^{-5} {\rm GeV^{-2}}$.}

Note that even though $\Cll_{1111}$ and $\Cll_{2222}$ do not contribute to R$(V_{us})$, 
it is quite difficult to come up with a high-scale theory
that can give rise to $\Cll_{1122}$ without also generating $\Cll_{1111}$ and $\Cll_{2222}$ at 
the same time. 

\section{The $Z'$ model}
\label{Zp}

The simplest extension of the SM that would give rise to nonzero 
$\Cll_{1122}(\Lambda)$ is the model with a heavy $Z'$ boson.
The Lagrangian of such a model may be written as
\begin{equation}
\mathcal{L}_{Z^\prime} = - g_{ij}^{\ell} \bar \ell_i \gamma^\mu \ell_j Z^\prime_\mu
- g_{ij}^{q} \bar q_i \gamma^\mu q_j   Z^\prime_\mu\,,
\end{equation}
where $i,j$ are fermion generation indices.
We take the leptonic couplings to be diagonal.
Since the off-diagonal leptonic couplings are severely constrained by 
the lepton-flavor violating (LFV) observables~\cite{Alok:2020jod},
postulating them to be vanishing would be a justified approximation.
This would allow all WCs of the form $\Cll_{iijj}$ to be nonzero
at the high scale $\Lambda$. However, this does not affect
eqs.~(\ref{eq:RVus-expansion})--(\ref{eq:RG-eps1}), so
our model-independent analysis above does not change.
Such a model will also not give rise to any $\Cphil, \Cphiq$, or
$\Clq$ WCs at the scale $\Lambda$.

On integrating out the heavy  $Z'$ boson, new dimension-six effective operators 
$\Oll_{iijj}$ and $\Olqone_{iimn}$, with
\begin{equation}
  \Olqone_{ijmn} = (\bar \ell_i \gamma^\mu \ell_j)
  (\bar q_m \gamma^\mu q_n) \,,
  \label{eq:Olq}
\end{equation}
are generated at the tree-level.
At the NP scale,  the WCs of these operators are
\begin{align}
  \Cll_{iijj}(\Lambda) & = -f \frac{g_{ii}^{\ell} g_{jj}^{\ell}} { M_{Z^\prime}^2} \; ,
  \label{eq:Cll-Z'} \\
\Clqone_{iimn}(\Lambda)  & = - \frac{g_{ii}^{\ell} g_{mn}^{q}}
       {M_{Z^\prime}^2}  \,,
       \label{eq:Clqone-Z'}
\end{align}
where $f=1/2$ for $i=j$ and $f=1$ otherwise. 
Note that WCs of the form $\Cll_{iiii}$ and $\Cll_{iijj}$ are related
through $(\Cll_{iijj})^2= 4~\Cll_{iiii} \cdot \Cll_{jjjj}$.
While nonzero $\Cll_{1122}(\Lambda)$ can help to resolve the CAA, 
nonzero $\Clqone_{2223}(\Lambda)$ can help us in resolving another
set of long-standing {$b\to s \mu^+ \mu^-$ anomalies}.

%%%%%%%%%%%%%%%%

The  current $b \to  s \mu^+ \mu^-$ data such as the branching ratio of $B_s\to \phi \mu^+ \mu^-$ and the optimized observable $P_5'$ 
exhibit some tension with the SM predictions \cite{bsphilhc2,bsphilhc3,LHCb:2020lmf,sm-angular}.
These can be accommodated by NP in the form of vector and axial-vector 
operators~\cite{Alok:2010zd,Alok:2011gv,Descotes-Genon:2013wba,Altmannshofer:2013foa,Hurth:2013ssa,Datta:2019zca,Kumar:2019qbv,Alok:2019ufo,
Altmannshofer:2021qrr,Carvunis:2021jga,Alguero:2021anc,
Geng:2021nhg,Hurth:2021nsi,Angelescu:2021lln,Ciuchini:2022wbq,SinghChundawat:2022ldm,SinghChundawat:2022zdf,Alguero:2023jeh}:
\begin{eqnarray}
  O_9^{bs\mu\mu} &\equiv&
  (\bar{s} \gamma^{\mu} P_L b) (\bar{\ell} \gamma^{\mu} \ell) \,, \\
  \label{eq:O9}
%%%
  O_{10}^{bs\mu\mu} &\equiv&
  (\bar{s} \gamma^{\mu} P_L b) (\bar{\ell} \gamma^{\mu} \gamma^5 \ell)~ .
  \label{eq:O10}
\end{eqnarray}
It is observed that one of the NP solutions preferred by the data is the one with the WCs
related by $C_9^{bs\mu\mu} = - C_{10}^{bs\mu\mu}$. 
In the context of the $Z'$ model, the operator $[O_{\ell q}^{(1)}]_{2223}$,
after the EW symmetry breaking, gives rise to the low-energy effective operators
$O_9^{bs\mu\mu}$ and $O_{10}^{bs\mu\mu}$ with 
\begin{equation}
  C_{9}^{bs\mu\mu}(\mu_{\rm EW}) = - C_{10}^{bs\mu\mu}(\mu_{\rm EW}) =
  \mathcal{N} \frac{[C_{\ell q}^{(1)}]_{2223}(\mu_{\rm EW})}{\Lambda^2}\, .
\end{equation}
In the basis used in {\tt flavio}  \cite{Straub:2018kue, Aebischer:2017ugx}, we have
 {$\mathcal {N} = \pi v^2/ (\alpha V_{tb} V_{ts}^*) $.}
The relation $C_{9}^{bs\mu\mu} = - C_{10}^{bs\mu\mu} $ is thus obtained
automatically \cite{Buras:2014fpa}.

%%%%%%%%%%%%%%%%%%%%%%%%%%%%%%%%%%%%%%%%%%%%%%%%%%%%%%%%%%%%%%

\section{Experimental constraints and fit results}
\label{results}

%\section{LFU-violating ratios $R_{K^{(*)}}$}
{The LFU is deeply embedded in the symmetry
structure of the SM.
The LHCb collaboration, in 2014, reported the measurement of the ratio
$R_K \equiv  \Gamma(B^+ \to K^+ \,\mu^+\,\mu^-)/
\Gamma(B^+ \to K^+\,e^+\,e^-)$ in the ``low-$q^2$'' range
($1.0\, {\rm GeV}^2 \le q^2 \le 6.0 \, {\rm GeV}^2$), where $q^2$ is the 
invariant mass-squared of the lepton pair~\cite{rk}.  
This measurement deviated from the SM value of $\simeq 1$ by 2.6$\sigma$,
and was the first strong indication of LFU violation in $b \to  s \ell^+ \ell^-$
decays. This was later corroborated by the measurement of the corresponding
ratio $R_{K^*}$ in $B^0 \to K^{*0} \ell^+\ell^-$ decays \cite{rkstar}.
In Moriond 2021, the LHCb collaboration reported an updated 
measurement of $R_K$ \cite{Aaij:2021vac} to be $0.846^{+0.044}_{-0.041}$. 
However, according to the latest LHCb update in 2022, these ratios are 
 measured to be consistent with the SM \cite{LHCb:2022zom}. 
Nevertheless, the $R_{K^{(*)}}$ remains an important measurement, whether for 
identifying LFU-violating new physics or for constraining the extent of LFU violation.}

%%%%%%%%%%%%%%%%%%%%%%%%%%%%%%%%%%%%%%%%%%%%%%%%%%%%%%%%%%%%%
\begin{figure}[th]
  \includegraphics[width=0.43\textwidth]{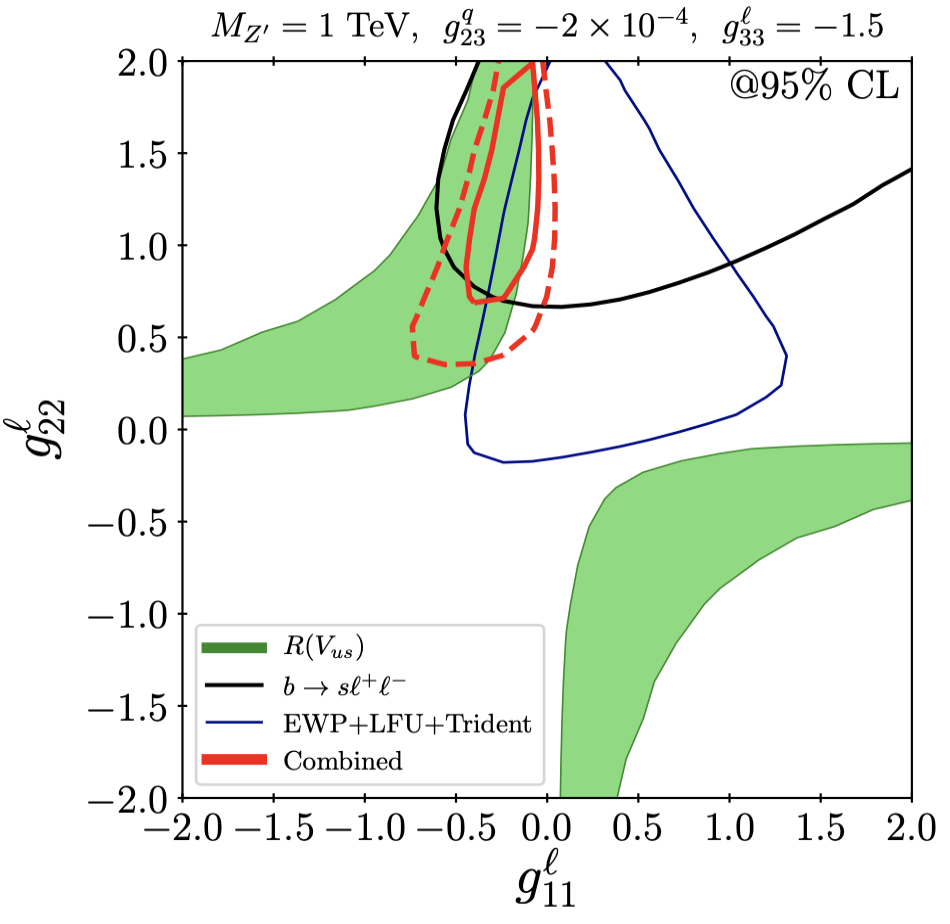}
  \includegraphics[width=0.45\textwidth]{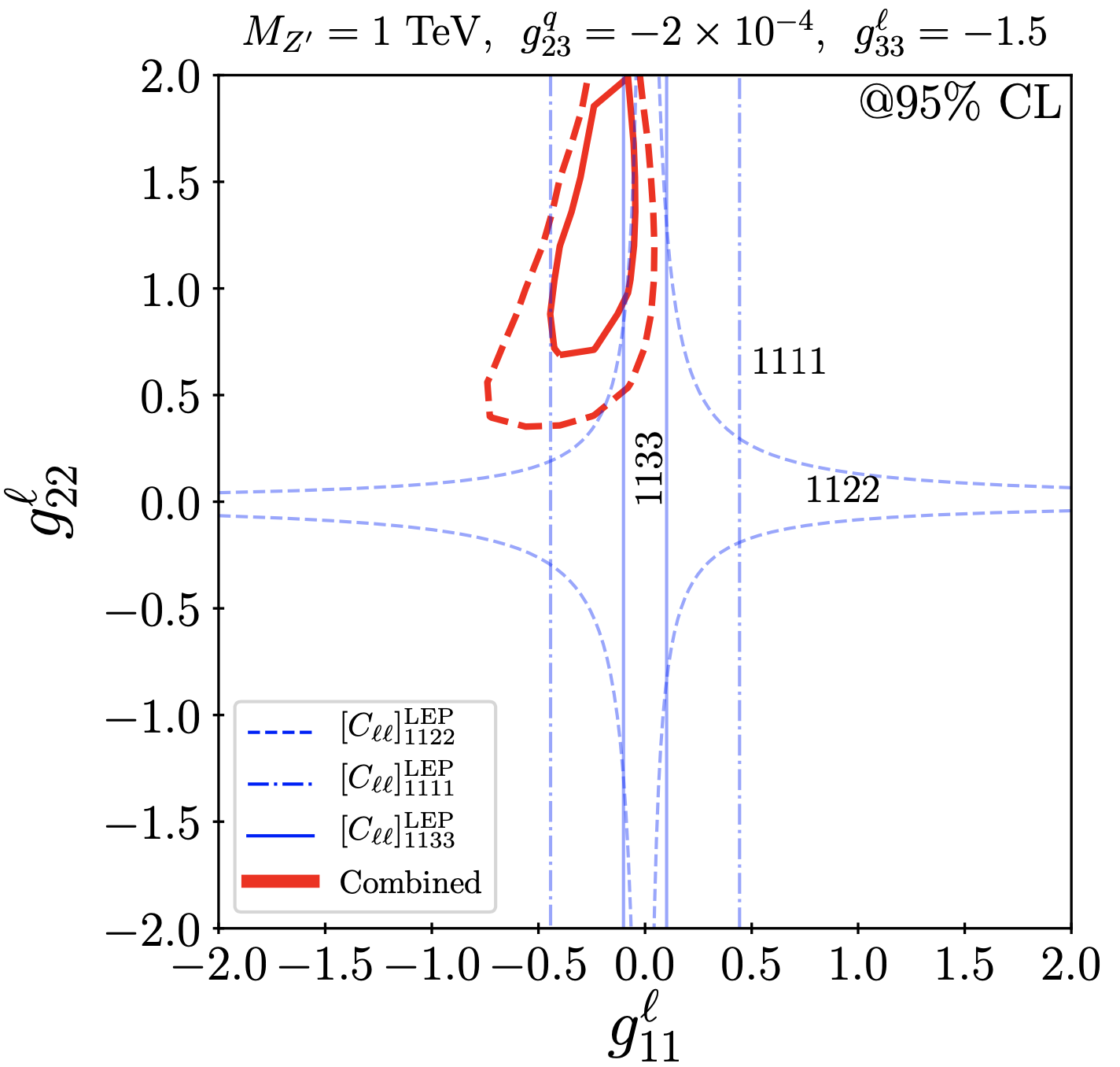}
  \caption{{In the left panel, we show regions in the $(g_{11}^\ell, g_{22}^\ell)$ 
parameter space indicated by the CAA anomaly (green). 
    Also shown are the bounds from the
    $b \to s \ell^+ \ell^-$ data (including $R_{K^{(*)}}$ 2023 LHCb update, black), 
    combination of EWP observables, LFU violating observables, and neutrino trident production (dark blue). 
    In the right panel, we show the LEP constraints (light blue) on the contact interactions. 
    All the regions correspond to $95 \%$ confidence level (C.L.), except the 
    combined fit ($R({V_{us}})$,  $b \to s \ell^+ \ell^-$, EWP observables, LFU-violating observables, 
neutrino trident) in red color, which is shown at $95\%$ C.L. (dashed) as well 
as at $68\%$ C.L. (solid) 
in both panels.
 }}
    \label{fig:ge-gmu}
\end{figure}
%%%%%%%%%%%%%%%%%%%%%%%%%%%%%%%%%%%%%%%%%%%%%%%%%%%%%%%%

The ATLAS and CMS collaborations have recently announced constraints on the
mass and couplings of the $Z'$ boson, based on its non-observation in the
di-muon channel, with $\sim 140 ~{\rm fb}^{-1}$ integrated luminosity in
each experiment~\cite{Aad:2019fac, CMS:2021ctt}.
Due to the smallness of the $bsZ^\prime$ coupling and the small fraction of 
$b$ and $s$ quarks inside the colliding protons, the data allow $M_{Z^{\prime}}$ 
values as low as a few hundred {GeV}~\cite{Allanach:2019mfl, Allanach:2018odd}.
However, we choose $M_{Z^{\prime}}=1$ TeV to ensure a cleaner separation of the
scale of NP from the EW scale, and hence, the validity of the EFT.
%{\purple This para may be shortened if space is needed for more material.}

The $Z^{\prime}$ model we consider is called the Mixed-Up Muon ‘MUM’ model 
(as defined in \cite{Allanach:2019mfl}), in which the $Z^{\prime}$ only couples to 
the $b$ and $s$ quarks 
and is produced via $\bar{s}b+\bar{b}s \to Z^{\prime}$  channel at the LHC. In this model, for $g_{bs}$ 
in the range [0.001 – 0.1], the constraint from $B_s$ mixing covers most of the 
region excluded by ATLAS dimuon searches (see fig. 4b of \cite{Allanach:2019mfl}). However, 
the value of $g_{bs}$ required to explain CAA and $b \to s \ell \ell$ anomalies is much 
smaller: $g_{bs} \sim 10^{-4}$, and for such small values there are currently no exclusion limits from ATLAS. 

The search capabilities of current and future experiments
 are highly model-dependent. For generic $g_{bs}$ couplings of $O(0.01)$, 
 for example, the projected sensitivity of the 3  $ {\rm ab}^{-1}$ HL-LHC to 
 the parameter space of the Mixed-Down Muon ‘MDM’ model is up to $M_{Z^{\prime}}$ = 5 TeV
  whereas it has no sensitivity to the MUM model \cite{CidVidal:2018eel}. 
  The proposed 27 TeV, 10 $ {\rm ab}^{-1}$  HE-LHC could probe $Z^{\prime}$ masses in the 
  MUM model up to 12 TeV.  The predicted sensitivity for this model at FCC 
  is up to $M_{Z^{\prime}}$ = 23 TeV \cite{Allanach:2018odd,FCC:2018byv}.

{
  We perform a global fit to $R(V_{us})$ and $b \to s\ell^+\ell^-$ observables including 
the latest measurements of  $R_{K^{(*)}}$, 
%data (144 observables) as in ,
  EWP observables (see \cite{Alok:2020jod} for the list of observables), % (27)
  LFU violating observables (see \cite{Alok:2020jod}), % (11), 
  and neutrino trident production in the
  $Z^\prime$ model, with $g_{11}^\ell$, $g_{22}^\ell$ and $g_{33}^\ell$
  as free parameters, keeping fixed values for 
  $M_{Z'}= 1$ TeV and $g_{23}^q = - 2 \times 10^{-4}$.
  Note that because of the relatively larger value of $g_{22}^{\ell}$
  {{required to account for CAA}}, the values of $g_{23}^{q}$ needed to accommodate
  the $b\to s \ell^+ \ell^-$ data are quite small. As a result, the constraints
  from $\Delta M_s$ are not significant. 
  We have employed {\tt flavio} and {\tt wilson} tools for the theoretical
  estimates of the observables and RG running, respectively.
  The fit yields
\begin{align}
  g_{11}^\ell  & =   -0.17\pm 0.10\,,  \qquad g_{22}^\ell =   +1.50 \pm 0.40\,, \qquad g_{33}^\ell  =    -1.80 \pm 0.90\,,
\end{align}
with $N_{\rm obs} = 135$, $\chi^2_{\rm SM} = 172.4$, and
$\chi^2_{\rm NP} \simeq 154.3$. The fit is thus a significant improvement over
the SM. At the best-fit point, we get $R(V_{us}) =0.9941$, which is well within 
 $1.5\sigma$ of the experimental value 
$0.9891 \pm 0.0033$ \cite{Seng:2018yzq}.

Our fit thus prefers a non-zero coupling of electrons as well as muons 
to $Z^\prime $. Further, a nonzero value of $g_{33}^\ell$ is needed to account for the 
 $\tau \to \mu \nu \bar \nu$ data. {The measured value of $A(\tau \to \mu \nu \nu)/ A(\mu \to e \nu \nu)$ is $1.0029 \pm 0.0014$ \cite{Amhis:2019ckw,ParticleDataGroup:2020ssz}, which differs from unity by about 2$\sigma$. Since $\mu \to e \nu \nu$ defines the “measured” Fermi constant, the explanation of the anomaly in the above ratio needs a non-zero value for $g_{33}^\ell$. The ratio is simply $1+ \epsphil_{33}- \epsphil_{11}$ , so no fine tuning is needed for this. }
Thus, $Z^\prime$ should couple to all 
three generations of the leptons. Note that it has also been argued
recently~\cite{Bhatia:2021eco} that
 $Z'$ couplings to all three flavors are needed in generic $Z'$ models
that address the $b \to s \ell^+ \ell^-$ anomalies and neutrino mixing pattern
simultaneously.

In fig.~\ref{fig:ge-gmu} (left panel), we show the region in the parameter space
of $(g_{11}^\ell, g_{22}^\ell)$ indicated by the data on $R(V_{us})$. 
It clearly prefers 
opposite signs for $g_{11}^\ell$ and $g_{22}^\ell$.
In $R(V_{us})$, this corresponds to positive $\Cll_{1122}$ [see eq.~(\ref{eq:Cll-Z'})].
The figure also shows the results of our separate fits to the
global $b \to s \ell^+ \ell^-$ data (including $R_{K^{(*)}}$),
and to the combined data from EWP observables,
LFU violating observables, and neutrino trident production~\cite{CHARM-II:1990dvf, CCFR:1991lpl}.
For $g_{22}^\ell > 0$, as strongly preferred by the
latter set of observables, a non-zero and negative $g_{11}^\ell$ is
needed to fit $R(V_{us})$. 
However, the global fit to the current $b\to s\ell^+ \ell^-$ data prefers the best fit 
in the first quadrant of  $(g_{11}^\ell, g_{22}^\ell)$ parameter space. 
This implies that  the future improvements in the $b\to s\ell^+\ell^-$ measurements 
have the potential to test the viability of our scenario.  

Note that the best-fit point preferred by our model is in tension with the LEP constraints
on the four-fermion contact interactions as obtained in \cite{ALEPH:2013dgf, Buras:2021btx}.
However, as can be seen in fig.~\ref{fig:ge-gmu} (right-panel), the $95\%$ C.L. 
allowed regions in the $(g_{11}^{\ell}, g_{22}^{\ell})$
plane allowed by all constraints do have an overlap with the LEP constraints.

Finally, it should be noted that in our fit we have used $m_W= 80.387 \pm 0.016 $ GeV.  
The recent CDF measurement of the $W$-mass \cite{CDF:2022hxs}, which is higher than the earlier $W$ mass 
measurements, has not been included.  
There have been attempts \cite{Cirigliano:2022qdm, Bagnaschi:2022whn,  Belfatto:2023tbv}  
to address this new anomaly in the 
SMEFT framework. These indicate that the value of $[C_{\ell\ell}]_{1221}$ 
(or equivalently $[C_{\ell\ell}]_{1122}$ at the high scale as used in our scenario) required to 
explain CAA decreases the value of $W$-mass  
as compared to the SM \cite{Bjorn:2016zlr}, and worsens the overall fit
\cite{Bagnaschi:2022whn}. 
Therefore, if the $W$-mass anomaly also has to be resolved along with the CAA and 
$B$ anomalies, then additional SMEFT operators would need to be invoked.

\section{Conclusions}
\label{summary}

In this article, we have proposed a new way to account for the CAA 
in the SMEFT framework, where we have used only purely leptonic 
operators at the high scale. We have shown that 
\begin{itemize} 

\item Pure leptonic four-fermion operators can affect the extraction of the  CKM 
element $V_{us}$ by contributing to the Fermi constant through operator mixing 
arising from RG evolution. 
The CAA, quantified through the ratio $R(V_{us})$, may be partly resolved 
  by the introduction of a single nonzero NP operator
  $\Oll_{1122}$ at a high scale $\Lambda$, and generating the required
  WCs at the low scale through RG running. The operators $\Oll_{1111}$
  and $\Oll_{2222}$ at the high scale do not contribute to the RG running
  of WCs relevant for the resolution of the CAA.

\item It is possible to generate nonzero values for
  $\Cll_{iijj}$  at the high scale, while
  keeping the WCs of other operators, $\Ophil, \Olq, \Ophiq$, to be
  vanishing at the high scale. This may be achieved, for example,
  through the extension of the SM with a heavy $Z'$ gauge boson 
having non-universal leptonic couplings.
In addition, in the $Z'$ model, the operator $\Olqone_{kk23}$ at the high scale
  can generate $C_9^{bs\ell \ell} = - C_{10}^{bs\ell\ell}$ at the EW scale,
  thus helping the resolution of the {$ b \to s \mu^+ \mu^-$} anomalies. 

\item Our model-independent scenario predicts that the value of $G_F^{(0)} \equiv 1/(\sqrt{2} v^2)$ in SMEFT
  is smaller than that in SM by  $\approx 0.05\%$, though the muon decay rate is the same. 
Therefore, it can be tested by precision measurements of the bare Fermi constant through CKM unitarity 
measurements and electroweak precision observables.
%Therefore, it 
%can be tested by precision measurements of the Fermi constant through other observables such as 
%muon decay and electroweak precision observables. 
Our scenario can also be tested by direct measurements of 
effective $ee\mu\mu$ coupling at future electron-positron collider such as FCC-ee or a muon collider.
In the context of the $Z'$ model, 
the desired values of $g_{11}^{\ell}$ and $g_{22}^{\ell}$ should be negative 
and positive, respectively. This prediction would be tested by precision 
measurements of $R_{K^{(*)}}$  in the future. 

\end{itemize}

%The current tensions in the CKM unitarity depend largely on forthcoming 
%precision calculations of nuclear corrections in nuclear beta decays and 
%advances on the experimental front possible through new measurements like BR()/BR() at the NA62 experiment.

The future of CAA hinges predominantly 
on the advancements in precision calculations of the nuclear 
corrections in beta decays. Moreover, progress on the experimental front, 
facilitated by measurements such as the ratio ${\rm BR}(K \to \pi \mu \bar \nu)/{\rm BR}(K \to \mu \bar \nu)$
possible at the NA62 experiment, would help to clarify if indeed the current tensions 
lead to unambiguous signals of NP. 
It will be exciting to see if the pattern of anomalies observed
in multiple channels at the low scale is actually pointing us to
a NP scenario at the high scale that is currently beyond the direct
search capabilities of particle colliders. 

\section{Acknowledgments} 
  The work of A.K.A. is supported by SERB-India Grant CRG/2020/004576.
  A.D. acknowledges support from the Department of Atomic Energy (DAE),
  Government of India, under Project Identification No. RTI4002. 
 The work of SG is supported by the ANR under contract n. 202650 (PRC `GammaRare').
 J.K. is financially supported by the Alexander von Humboldt Foundation's postdoctoral research
fellowship. J.K. thanks Christoph Bobeth for useful discussions. 
{ We also thank Andreas Crivellin and Teppei Kitahara for
incisive comments on an earlier version of this paper.}

\nocite{*}


\end{document}